\def   \ni {\noindent}

\def   \ssk {\vskip  5truept}

\def   \bsk {\vskip 15truept}
 
\def   \newpage {\vfill\eject}
\def   \newline {\hfil\break}

\def   \sssk {\vskip  1truept}

\documentstyle[psfig]{article}
\begin{document}

\hsize 5truein
\vsize 8truein
\font\abstract=cmr8
\font\keywords=cmr8
\font\caption=cmr8
\font\references=cmr8
\font\text=cmr10
\font\affiliation=cmssi10
\font\author=cmss10
\font\mc=cmss8
\font\title=cmssbx10 scaled\magstep2
\font\alcit=cmti7 scaled\magstephalf
\font\alcin=cmr6 
\font\ita=cmti8
\font\mma=cmr8
\def\ref{\par\noindent\hangindent 15pt}
\null


\title{\ni COMPTEL DETECTION OF PULSED EMISSION FROM PSR B1509-58 UP TO AT LEAST 
10 MEV}

\bsk 
\bsk
\author{\ni L.~Kuiper$^{1}$, W.~Hermsen$^{1}$, J.M.~Krijger$^{1}$, K.~Bennett$^{2}$, 
V. Sch\"onfelder$^{3}$, A.~Carrami\~nana$^{4}$, R.~Manchester$^{5}$, M.~Bailes$^{6}$}

\bsk
\affiliation{1) SRON-Utrecht, Sorbonnelaan 2, 3584CA Utrecht, The Netherlands}
\sssk
\affiliation{2) Astrophysics Division, ESTEC, 2200AG Noordwijk, The Netherlands}
\sssk
\affiliation{3) Max-Planck Institut f\"ur Extraterrestrische Physik, D-8046 Garching, Germany}
\sssk
\affiliation{4) INAOE, Luis Enrique Erro 1, Tonantzintla, Puebla 72840, Puebla, M\'exico}
\sssk
\affiliation{5) Australian Telescope National Facility, CSIRO, Epping, Australia}
\sssk
\affiliation{6) Physics Department, University of Melbourne, Australia}
\bsk
\baselineskip = 12pt

\abstract{ABSTRACT \ni We report the COMPTEL detection of pulsed $\gamma$-emission from PSR B1509-58
up to at least 10 MeV using data collected over more than 6 years. The 0.75-10 MeV lightcurve is broad
and reaches its maximum near radio-phase 0.38, slightly beyond the maximum found at hard X-rays/ 
soft $\gamma$-rays. In the 10-30 MeV energy range a strong source is present in the skymap positionally 
consistent with the pulsar, but we do not detect significant pulsed emission. However, the lightcurve is
consistent with the pulse shape changing from a single broad pulse into a double-peak morphology. 
Our results significantly constrain pulsar modelling.}                                                    
\bsk
\baselineskip = 12pt

\bsk
\baselineskip = 12pt



\text{\ni 1. INTRODUCTION
\ssk
\ni
     
PSR B1509-58 was discovered in the late seventies as a 150 ms X-ray pulsar in Einstein data of the 
Supernova Remnant MSH 15-52 (Seward et al. 1982). Its inferred characteristic age 
and polar surface magnetic field strength are 1570 year and $3.1\times 10^{13}$ Gau\ss.
The latter estimate is among the highest of the radio-pulsar population.
Ginga (2-60 keV) detected pulsed emission at hard X-rays (Kawai et al. 1991), the profile being broad and 
asymmetric, and its maximum trails the radio-pulse $0.25\pm0.02$ in phase. 
After the launch of the Compton Gamma-Ray Observatory (CGRO) pulsed emission in the soft $\gamma$-ray 
band was seen by BATSE (Wilson et al. 1993) and OSSE (Ulmer et al. 1993; Matz et al. 1995).
At medium energy $\gamma$-rays indications were found near $\sim 1$ MeV in COMPTEL data (Hermsen et al.
1994; Carrami\~nana et al. 1995). 
The non-detection by EGRET (e.g. Brazier et al. 1994) indicates that the pulsed spectrum must break before
the high-energy (HE) $\gamma$-rays.
Here we report the results from a COMPTEL (0.75-30 MeV) study of PSR B1509-58 using data collected over
more than 6 years, applying improved event selections. More detailed information will be given in Kuiper et al. (1999).

\newpage

\bsk
\ni 2. COMPTEL TIMING ANALYSIS RESULTS
\ssk
\ni

The arrival times of the selected events at the spacecraft, each recorded with a 0.125 ms accuracy,
have been converted to Solar System Barycentric arrival times and subsequently phase folded with a
proper radio-pulsar ephemeris. The resulting 0.75-30 MeV lightcurve shows a $5.4\sigma$ modulation 
significance ($Z_2^2$-test) with a single pulse roughly aligned with the pulse detected at lower energies.
The maximum of the pulse is at phase 0.38 slightly above 0.27 found at hard X-rays and coincides with the ``shoulder''
in the RXTE 2-16 keV lightcurve (Fig. 1a). For the differential energy windows 0.75-3 MeV, 3-10 MeV
and 10-30 MeV we show the lightcurves in Fig. 1; the modulation significances are $3.7\sigma, 4.0\sigma$ and $2.1\sigma$,
respectively. This proves that we {\em detected} pulsed emission from this source at least up to 10 MeV.
The 10-30 MeV lightcurve (Fig. 1b) shows an indication for the broad pulse and a high bin near phase 0.85, which seems
to be responsible for the low modulation significance. 
Based on the RXTE lightcurve (Fig. 1a) we defined an ``unpulsed'' phase interval: 0.65-1.15. 
The excess counts in the ``pulsed'' interval, 0.15-0.65, have been converted to flux estimates
for the 3 COMPTEL energy windows taking into account the efficiency factors due to our event selection 
criteria. These ``pulsed'' fluxes are: $(3.69\pm0.73)\times 10^{-5}$, $(4.52\pm0.77)\times 10^{-6}$ and 
$(1.21\pm0.85)\times 10^{-7}$ in units $ph / cm^2\cdot s\cdot MeV$ for the energy windows 0.75-3 MeV, 
3-10 MeV and 10-30 MeV, respectively.

\bsk
\ni 3. COMPTEL IMAGING ANALYSIS RESULTS
\ssk
\ni
 
We performed imaging analyses selecting also on pulse phase (``Total'', phase range 0-1; ``Pulsed'' and ``Unpulsed''
intervals). Below 10 MeV the signal from PSR B1509-58 is consistent with being 100\% pulsed (Kuiper 
et al. 1999). In the 10-30 MeV range, where the timing analysis did not reveal significant modulation, we 
surprisingly detect in the ``Total'' map a $\sim 6\sigma$ source atop the instrumental and galactic diffuse 
background positionally consistent with the pulsar (see Collmar et al., these proceedings). In view of the absence of any 
measurable pulsar/nebula DC emission or nearby unrelated source below 10 MeV, where COMPTEL is more sensitive, we 
consider it most likely that the pulsar is also detected above 10 MeV, possibly changing its pulse morphology.
In order to find support for the latter interpretation we also analysed contemporaneous
EGRET 30-100 MeV data.

\begin{figure}[t]


  \centerline{
    \parbox[b]{0.5\columnwidth}
    {\psfig{file=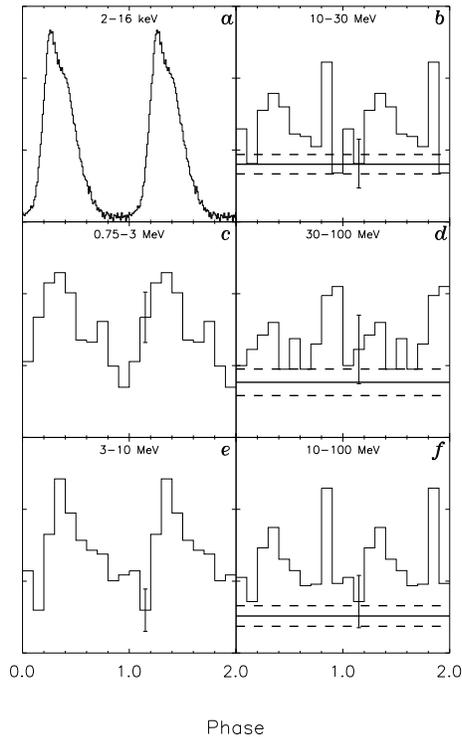,height=10.75cm}}
    \parbox[b]{0.425\columnwidth}
    {\caption{
    FIGURE 1. Radio-aligned lightcurves of PSR B1509-58: {\bf a)} RXTE 2-16 keV
    (Rots et al. 1998) {\bf b)} COMPTEL 10-30 MeV (2.1$\sigma$) {\bf c)} COMPTEL 0.75-3 MeV 
    (3.7$\sigma$) {\bf d)} EGRET 30-100 MeV (1.9$\sigma$) {\bf e)} COMPTEL 3-10 MeV
    (4.1$\sigma$) and {\bf f)} COMPTEL \& EGRET 10-100 MeV (2.3$\sigma$). Typical
    error bars are indicated. In Figs. b, d, f background levels and 1$\sigma$ errors determined in the 
    imaging analysis are indicated as solid and dashed lines.\vspace{3.5truecm}
             }
    }}
\end{figure}

\bsk
\ni 4. EGRET 30-100 MeV AND COMPTEL 10-30 MeV RESULTS
\ssk
\ni

In the spatial analysis in the EGRET 30-100 MeV energy window we detected a $ 6.7\ \sigma$ excess positionally consistent
with PSR B1509-58, which is probably composed of contributions from both the unidentified EGRET source 
2EG J1443-6040 and PSR B1509-58. In the timing analysis, applying the same event selections as in the spatial analysis 
in combination with a ``standard'' energy dependent cone selection (Thompson et al. 1996), we found only a $1.1 \sigma$ deviation from 
a flat distribution (Fig. 1d). Summing the independent COMPTEL 10-30 MeV and EGRET 30-100 MeV lightcurves results in a 
suggestive double-peak lightcurve with a marginally significant modulation of $2.3\sigma$ (Fig. 1f). If the 
double-peak interpretation is correct, we have underestimated our 10-30 MeV ``pulsed'' ( phase 0.15-0.65) flux in the timing 
analysis (Sect. 2) and then the flux of the broad pulse increases to $(3.37\pm0.70)\times 10^{-7}\ ph / cm^2\cdot s\cdot MeV$. 
Our flux estimates (0.75-30 MeV) are shown in Fig. 2 along with spectra from other HE-instruments.  
It is clear from this picture that the ``pulsed'' spectrum breaks above 10 MeV.
}

\bsk
\ni 5. COMPARISON WITH THEORY 
\ssk
\ni

\begin{figure}[t]


  \centerline{
    \parbox[b]{0.5\columnwidth}
    {\psfig{file=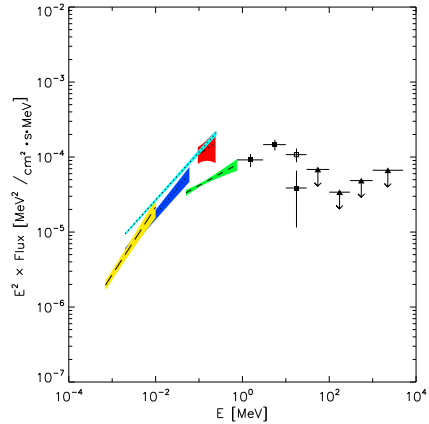,height=6truecm}}
    \parbox[b]{0.48\columnwidth}
    {\caption{FIGURE 2. Pulsed spectrum of PSR B1509-58 from soft X-rays up to
     HE- $\gamma$-rays: yellow polygon ASCA (Saito et al. 1997); dark blue Ginga (Kawai et al. 1993);
     light blue RXTE (Marsden et al. 1998); red Welcome (Gunji et al. 1994); green OSSE (Matz et al. 1995), 
     filled squares COMPTEL (this work; 10-30 MeV 2 flux estimates (see text)) and filled triangles EGRET
     (Thompson, D.; private communication)}\vspace{0.65truecm}}
             }

\end{figure}


For explaining pulsed HE-radiation two catagories of models exist, polar cap and outer gap, with 
as main difference the production site of the HE-radiation.
Recently, in a polar cap scenario Harding et al. (1997) (see also Baring \& Harding these proceedings) tried to 
explain the earlier HE-spectrum of PSR B1509-58 by including the exotic photon splitting process, only effective 
for magnetic field strengths near $\sim 4.41\times 10^{13}$ Gau\ss,~to attenuate the primary HE-$\gamma$-rays in 
their cascade calculations. Our new 0.75-30 MeV data  constrain the magnetic co-latitude of the emission rim, one 
of their model parameters, to be about $2^{\circ}$, close to the ``classical'' polar cap half-angle.
In the outer gap scenario (Romani 1996) several distinct HE-radiation components can be identified.
If the synchrotron component, most important at medium energy $\gamma$-rays, is dominant, then also this
model might qualitatively explain the HE-spectrum of PSR B1509-58. Another high B-field young pulsar resembling 
PSR B1509-58, although $\sim 20$ times weaker in spin-down flux, is PSR B1610-50. This pulsar will also be a 
promising target for future INTEGRAL observations.


\bsk
\baselineskip = 12pt


{\references \ni REFERENCES
\ssk
\ref Brazier, K. T. S, Bertsch, D. L., Fichtel, C. E., et al., 1994, MNRAS 268, 517
\ref Carrami\~nana, A., Bennett, K., Buccheri, R., et al., 1995, A\&A 304, 258
\ref Gunji, S., Hirayama, M., Kamae, T., et al., 1994, ApJ 428, 284
\ref Harding, A., Baring, M., Gonthier, P., 1997, ApJ 476, 246
\ref Hermsen, W., Kuiper, L., Diehl, R., et al., 1994, ApJS 92, 559
\ref Kawai, N., Okayasu, R., Brinkmann, W., et al., 1991, ApJ 383, L65
\ref Kawai, N., Okayasu, R., Sekimoto, Y., 1993, AIP Conf. Proc.  280, p. 213
\ref Kuiper, L., Hermsen, W., Krijger, J., et al., 1999, A\&A (in preparation)
\ref Marsden, D., Blanco, P. R., Gruber, D. E., et al. 1998, ApJ 491, L39
\ref Matz, S. M., Ulmer, M. P., Grabelsky, D. A., et al. 1995, ApJ 434, 288
\ref Romani, R., 1996, ApJ 470, 469                     
\ref Rots, A. H., Jahoda, K., Macomb, D. J., et al., 1998, ApJ 501, 749                     
\ref Saito, Y., Kawai, N., Kamae, T., et al., 1997, AIP Conf. Proc. 410, p. 628
\ref Seward, F. D., Harnden Jr., F. R., 1982, ApJ 256, L45                     
\ref Thompson, D. J., Bailes, M., Bertsch, D. L., et al., 1996, ApJ 465, 385                     
\ref Ulmer, M. P., Matz, S. M., Wilson, R. B., et al., 1993, ApJ 417, 738
\ref Wilson, R.B., Fishman, G. J., Finger, M. H., et al., 1993, AIP Conf.
Proc.  280, p. 291
}

\end{document}